\documentclass[twocolumn,floatfix,aps,superscriptaddress]{revtex4}
\usepackage{graphicx}
\usepackage{amsmath}
\usepackage{amssymb}
\usepackage{bm}
\usepackage{hyperref}

\begin{document}

\title{Scattering theory of the chiral magnetic effect in a Weyl semimetal:\\ 
Interplay of bulk Weyl cones and surface Fermi arcs}
\author{P. Baireuther}
\affiliation{Instituut-Lorentz, Universiteit Leiden, P.O. Box 9506, 2300 RA Leiden, The Netherlands}
\author{J. A. Hutasoit}
\affiliation{Instituut-Lorentz, Universiteit Leiden, P.O. Box 9506, 2300 RA Leiden, The Netherlands}
\author{J. Tworzyd{\l}o}
\affiliation{Institute of Theoretical Physics, Faculty of Physics, University of Warsaw, ul.\ Pasteura 5, 02--093 Warszawa, Poland}
\author{C. W. J. Beenakker}
\affiliation{Instituut-Lorentz, Universiteit Leiden, P.O. Box 9506, 2300 RA Leiden, The Netherlands}
\date{December 2015}
\begin{abstract}
We formulate a linear response theory of the chiral magnetic effect in a finite Weyl semimetal, expressing the electrical current density $j$ induced by a slowly oscillating magnetic field $B$ or chiral chemical potential $\mu$ in terms of the scattering matrix of Weyl fermions at the Fermi level. Surface conduction can be neglected in the infinite-system limit for $\delta j/\delta \mu$, but not for $\delta j/\delta B$: The chirally circulating surface Fermi arcs give a comparable contribution to the bulk Weyl cones no matter how large the system is, because their smaller number is compensated by an increased flux sensitivity. The Fermi arc contribution to $\mu^{-1}\delta j/\delta B$ has the universal value $(e/h)^2$, protected by chirality against impurity scattering --- unlike the bulk contribution of opposite sign. 
\end{abstract}
\maketitle

\section{Introduction}
 
The conduction electrons in a Weyl semimetal have an unusual velocity distribution in the Brillouin zone \cite{Vol03}. The conical band structure (Weyl cone) has a chirality that generates a net current at the Fermi level in the presence of a magnetic field \cite{Nie83}. The Weyl cones come in pairs of opposite chirality, so that the total current vanishes in equilibrium \cite{Vaz13,Zho13,Bas13}, but a nonzero current $I$ parallel to the field $B$ remains if the cones are offset by an energy $\mu$ --- slowly oscillating to prevent equilibration \cite{Che13,Gos13,Cha14,Ala15,Ma15,Zho15}. This is the chiral magnetic effect (CME) from particle physics \cite{Vil80,Ale98,Gio98,Fuk08}, see Refs.\ \onlinecite{Hos13,Kha14,Bur15} for recent reviews in the condensed matter setting. In an infinite system the current density has the universal form \cite{Zyu12a,Zyu12b,note0}
\begin{equation}
j_0=-(e/h)^2 \mu B ,\label{jCME}
\end{equation}
independent of material parameters. This amounts to a conductance of $e^2/h$ in the lowest (zeroth) Landau level, multiplied by the degeneracy equal to the enclosed flux in units of the flux quantum.

The recent condensed-matter realizations of Weyl semimetals \cite{Xu15,Lv15,Yan15,NXu15} have boosted the search for the chiral magnetic effect \cite{Xio15,Hua15,Li14,Zha15,CZha15,Beh15,She15,Kha15,reviews}. Future experimental developments may well include nanostructured materials, to minimize effects of disorder. In a finite system, the zeroth Landau level in the bulk hybridizes with the Fermi arcs connecting the two Weyl cones along the surface \cite{Wan11,Hal14}. Previous studies \cite{Gor15,Val15} have pointed to the importance of boundaries for the chiral magnetic effect --- a sign reversal of the current density as one moves from the bulk towards a boundary ensures that zero current flows in response to a static perturbation. Here we wish to study how this interplay of surface and bulk states impacts on the chiral magnetic effect in response to a low-frequency dynamical perturbation. For that purpose we seek a linear response theory that does not assume translational invariance in an infinite system. A scattering formulation \textit{\`{a} la} Landauer seems most appropriate for such a mesoscopic system.

The Landauer approach to electrical conduction considers the current driven between two spatially separated electron reservoirs by a chemical potential difference, and expresses this in linear response by a sum over transmission probabilities at the Fermi level \cite{Dat97,Imr08,Naz09}. The chiral magnetic effect is driven by a nonequilibrium population of the Weyl cones, so in reciprocal space (Brillouin zone) rather than in real space --- we will show how to modify the Landauer formula accordingly. 

We first apply our scattering formula to a current driven by a slowly oscillating offset $\mu$ of the Weyl cones (a so-called ``chiral'' or ``axial'' chemical potential \cite{Fuk08}), and recover Eq.\ \eqref{jCME} in the infinite-system limit. We then turn to the more practical scenario of a current driven by a slowly oscillating magnetic field $B$. We find that the surface Fermi arcs give a contribution to the total induced current equal to {\em minus twice\/} the bulk contribution in the infinite-system limit. That the surface Fermi arc contribution does not vanish relative to the bulk contribution is unexpected and not captured by previous calculations of the chiral magnetic effect.

The outline of the paper is as follows. In the next section \ref{sec_scattering} we derive the scattering formula for the chiral magnetic effect, in a general setting. In Sec.\ \ref{sec_induced} we apply it to the model Hamiltonian of a Weyl semimetal from Ref.\ \onlinecite{Vaz13}, summarized in Sec.\ \ref{sec_model}. We evaluate the induced current in response to variations in $\mu$ and $B$, both numerically for a finite system and analytically in the limit of an infinite system size. Finite-size corrections are considered in some detail in Sec.\ \ref{sec_finite}. We conclude in Sec.\ \ref{sec_conclude} with a summary and a discussion of the robustness of the results against disorder scattering.

\section{Scattering formula}
\label{sec_scattering}
 
For a scattering theory of the chiral magnetic effect we consider a disordered mesoscopic system attached to ideal leads. Such an ``electron wave guide'' has propagating modes with band structure $E_n(k)$, labeled by a mode index $n=1,2,\ldots$ and dependent on the wave vector $k$ along the lead. At a given energy $\varepsilon$ (measured relative to the equilibrium Fermi level $E_{\rm F}$), each incident mode has wave vector $k_n(\varepsilon)$ and carries the same current $e/h$ per unit energy interval \cite{note1}.

The scattering matrix $S(\varepsilon)$ relates amplitudes of incident and outgoing modes. We take a two-terminal geometry (the multi-terminal generalization is straightforward), with $N$ modes each in the left and right lead --- so $S$ is a $2N\times 2N$ unitary matrix. The current $I$ through the system can be calculated in the left lead, by current conservation it must be the same through each cross section. 

The projection matrix onto the left lead is ${\cal P}={{1\,0}\choose{0\,0}}$, where each sub-block is an $N\times N$ matrix. The current is driven by a set of non-equilibrium occupation numbers $\delta f_n(\varepsilon)$, with $n=1,2,\ldots N$ for the left lead and $n=N+1,N+2,\ldots 2N$ for the right lead. We collect these numbers in a $2N\times 2N$ diagonal matrix $\delta {\cal F}(\varepsilon)$. The net current in the left lead is then given by the difference of incoming and outgoing currents,
\begin{equation}
I=\frac{e}{h}\int d\varepsilon\,{\rm Tr}\,\left[{\cal P}\delta{\cal F}(\varepsilon)-{\cal P}S(\varepsilon)\delta{\cal F}(\varepsilon)S^\dagger(\varepsilon)\right].\label{IDeltaF}
\end{equation}

We consider the linear response to a slowly varying parameter $X$ that adiabatically perturbs the system away from its equilibrium state at $X=X_0$. We assume that the wave vector $k$ along the lead (say, in the $z$-direction) is not changed by the perturbation. This requires that the perturbation should neither break the translational invariance along $z$, nor involve a time-dependent vector potential component $A_z$.

The band structure evolves from $E_n(k|X_0)$ to $E_n(k|X_0+\delta X)$. To first order in the perturbation $\delta X$ the energy shift at constant $k$ is
\begin{equation}
E_n(k|X_0+\delta X)-E_n(k|X_0)=\delta X\lim_{X\rightarrow X_0}\frac{\partial}{\partial X} E_n(k|X).\label{deltaElinear}
\end{equation}
The corresponding deviation of the occupation number from the equilibrium Fermi function $f_{\rm eq}(\varepsilon)=(1+e^{\varepsilon/k_{\rm B}T})^{-1}$ is
\begin{align}
\delta f_n(\varepsilon)&=f_{\rm eq}\biglb(E_n(k_n(\varepsilon)|X_0)\bigrb)-f_{\rm eq}(\varepsilon)\nonumber\\
&=-\delta X f'_{\rm eq}(\varepsilon)\lim_{X\rightarrow X_0}\frac{\partial}{\partial X}E_n(k_n(\varepsilon)|X),\label{deltaflinear}
\end{align}
where we have used that $E_n(k_n(\varepsilon)|X_0+\delta X)\equiv\varepsilon$.

At zero temperature the derivative $f'_{\rm eq}(\varepsilon)\rightarrow -\delta(\varepsilon)$, so the expression \eqref{IDeltaF} for the current contains only Fermi level scattering amplitudes. We may write it in a more explicit form in terms of the transmission probabilities
\begin{equation}
T_n=\begin{cases}
\sum_{m=N+1}^{2N}|S_{mn}|^2&{\rm for}\;\;1\leq n\leq N,\\
\sum_{m=1}^{N}|S_{mn}|^2&{\rm for}\;\;N+1\leq n\leq 2N,
\end{cases}
\end{equation}
evaluated at $\varepsilon=0$. (The two cases correspond to transmission from left to right or from right to left.) Since $\sum_{m=1}^{2N}|S_{mn}|^2=1$ because of unitarity, we have
\begin{equation}
\begin{split}
&I=\frac{e}{h}\delta X\sum_{n=1}^{2N}\chi_n T_n,\\
&\chi_n=\lim_{k\rightarrow k_n}\lim_{X\rightarrow X_0}\frac{\partial E_n(k|X)}{\partial X}\times {\rm sign}\,\frac{\partial E_n(k|X)}{\partial k}.
\end{split}\label{IzeroT}
\end{equation}
The sign of the derivative $\partial E_n/\partial k$ distinguishes the right-moving modes $n=1,2,\ldots N$ from the left-moving modes $n=N+1,N+2,\ldots 2N$.

The Landauer conductance formula \cite{Dat97,Imr08,Naz09}
\begin{equation}
G=I/V=(e^2/h)\textstyle{\sum_{n=1}^N} T_n\label{LandauerG}
\end{equation}
is obtained from Eq.\ \eqref{IzeroT} if we identify $\delta X=V$ with the voltage difference between the left and right lead, and then set $\chi_n=1$ for $n=1,2,\ldots N$ and $\chi_n=0$ for $n=N+1,N+2,\ldots 2N$. The chiral magnetic effect is driven by a non-equilibrium population in momentum space, rather than in real space, so modes from both leads contribute --- hence the need to sum over $2N$ rather than $N$ modes.

\section{Model Hamiltonian of a Weyl semimetal}
\label{sec_model}

A simple model of a Weyl semimetal is given by the four-band Hamiltonian \cite{Vaz13}
\begin{align}
H(\bm{k})={}&t'\tau_z(\sigma_x\sin k_x+\sigma_y\sin k_y) +t'_z \tau_y\sin k_z\nonumber\\
&+M(\bm{k})\tau_x+\tfrac{1}{2}\gamma\tau_y\sigma_z+\tfrac{1}{2}\beta\sigma_z,\label{HMdef}\\
M(\bm{k})={}&M_0+t(2-\cos k_x-\cos k_y)+t_z(1-\cos k_z).\nonumber
\end{align}
The Pauli matrices $\sigma_j$ and $\tau_j$ ($j=x,y,z$) act, respectively, on the spin and orbital degree of freedom. The momentum $\bm{k}$ varies over the Brillouin zone $-\pi<k_j<\pi$ of a simple cubic lattice (lattice constant $a\equiv 1$). The material is layered in the $x$-$y$ plane, with nearest-neighbor hopping energies $t$ (within the layer) and $t_z$ (along the $z$-axis). The primed terms $t',t'_z$ indicate hopping with spin-orbit coupling. Inversion symmetry, $\tau_x H(-\bm{k})\tau_x=H(\bm{k})$, is broken by strain $\propto\gamma$, while time-reversal symmetry, $\sigma_y H^\ast(-\bm{k})\sigma_y=H(\bm{k})$, is intrinsically broken by a magnetization $\propto\beta$. Additionally, we may apply a magnetic field in the $+z$-direction, by substituting $k_y\mapsto k_y-eBx/\hbar$. The field strength is characterized by the magnetic length $l_B=\sqrt{\hbar/eB}$.

\begin{figure}[tb]
\centerline{\includegraphics[width=1\linewidth]{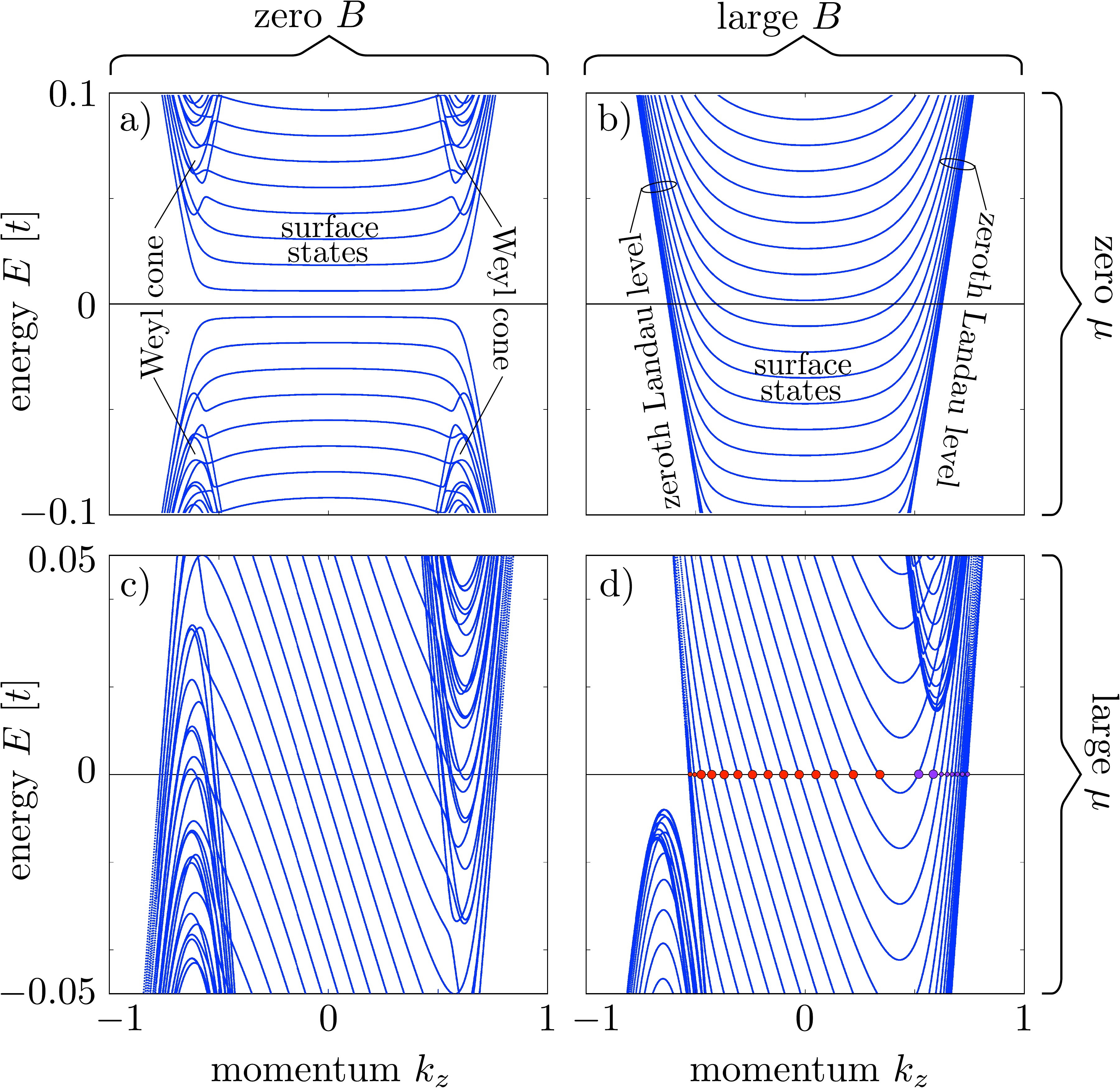}}
\caption{Band structure of the Hamiltonian \eqref{HMdef} in a wire geometry along the $z$-axis. The
  panels show the Weyl cones in zero magnetic field (panels a,c), the
  Landau levels in a strong magnetic field along $z$ (panels b,d for
  $l_{B}=25$), each for $\pm k_z$ symmetry (panels a,b) and when this
  inversion symmetry is broken (panels c,d for $\gamma=0.2\,t\Rightarrow\mu=0.196\,t$). The
  other model parameters are $t_z=t'_z=t$, $t'=2t$, $\beta=1.2\,t$,
  $M_0=-0.3\,t$, $W_x=W_y=255$ in units of the lattice constant
  $a$. The intersections of the subbands $E_n(k_z)$ with the Fermi
  level $E_{\rm F}=0$ determine the momenta $k_n$ appearing in the
  scattering formula \eqref{IzeroT}. These are indicated by dots in
  panel d, colored purple or red depending on whether the mode
  propagates in the $+z$ or in the $-z$ direction (as determined by
  the sign of $dE_n/dk_z$).  }
\label{fig_bandstructure}
\end{figure}

\begin{figure}[tb]
\centerline{\includegraphics[width=0.4\linewidth]{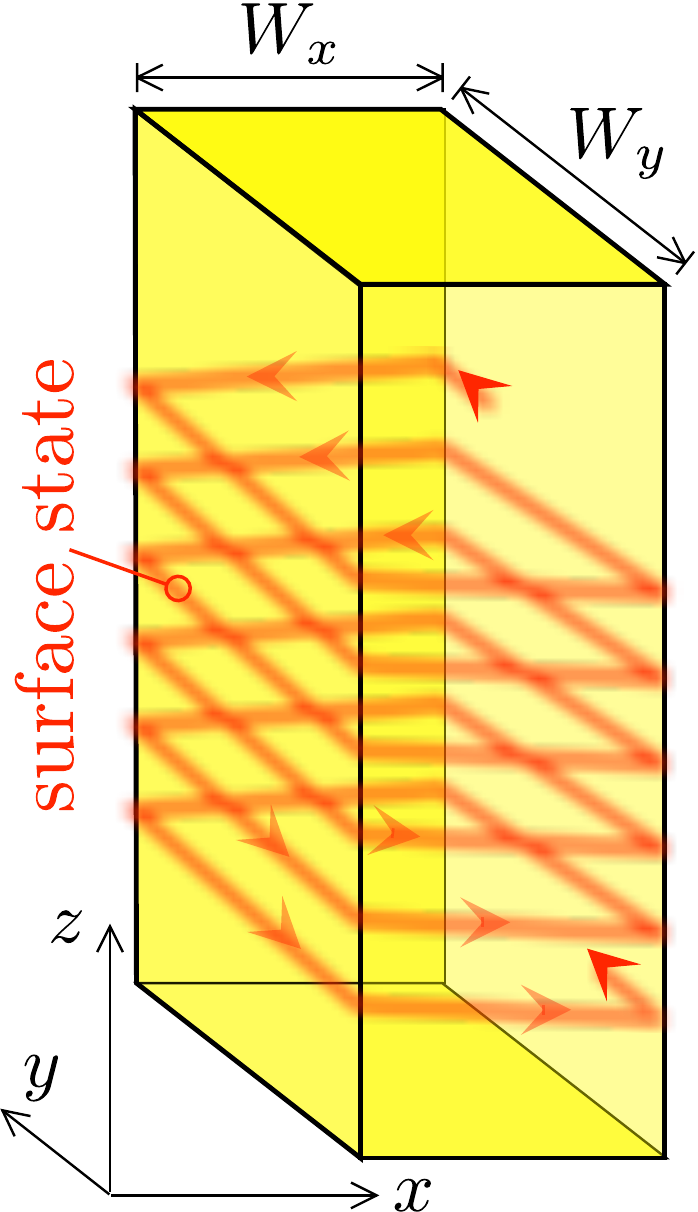}}
\caption{\textit{Weyl solenoid}. Illustration of a chiral surface Fermi arc spiraling along the wire (cross-sectional area ${\cal A}$ and perimeter ${\cal P}$). Its flux sensitivity is set by the orbital magnetic moment $ev_{\rm F} {\cal A}/{\cal P}$, while the number of surface modes at the Fermi level scales $\propto {\cal P}$, so their total contribution to the magnetic response is $\propto{\cal A}$ --- of the same order as the bulk contribution.
}
\label{fig_spiral}
\end{figure}

We confine the layers to a $W_x\times W_y$ lattice in the $x$-$y$
plane, infinite in the $z$-direction so $k_z$ remains a good quantum
number. The tight-binding Hamiltonian in this wire geometry is
diagonalized with the help of the {\sc kwant} toolbox \cite{kwant},
see Fig.\ \ref{fig_bandstructure}. In zero magnetic field (panels a,c)
there are two Weyl cones, gapped by the finite system size. The conical
points (Weyl points) are separated along $k_z$
by approximately $\beta/t'_z$ and they are separated in energy by approximately $\gamma$.
The precise energy separation $\mu$ that governs the chiral magnetic effect was determined 
from the bandstructure in an infinite system, for our parameter values it differs from $\gamma$ by a few percent.

As long as $\mu,M_0\ll\beta$ the Weyl cones remain
distinct in an energy interval around $E=0$. The Fermi velocity of the
massless Weyl fermions is $v_{\rm F}=t'/\hbar$ in the plane of the
layers and $v_{{\rm F},z}=t'_z/\hbar$ perpendicular to the layers.
Surface states connect the Weyl cones across the Brillouin zone, forming the so-called Fermi arc. The arc states are chiral, spiraling along the wire with a velocity $v_{{\rm arc},z}=(\mu/\beta)v_{{\rm F},z}$, as illustrated in Fig.\ \ref{fig_spiral}. In a magnetic field (panels b,d in Fig.\ \ref{fig_bandstructure}) Landau levels develop. The Weyl cones are pushed away from $E=0$, but the zeroth Landau level closes the gap. Just like the Fermi arc, the zeroth Landau level propagates along the wire, in opposite direction for the two Weyl cones.

\section{Induced current in linear response}
\label{sec_induced}

\subsection{Numerical results from the scattering formula}

\begin{figure}[tb]
\centerline{\includegraphics[width=0.8\linewidth]{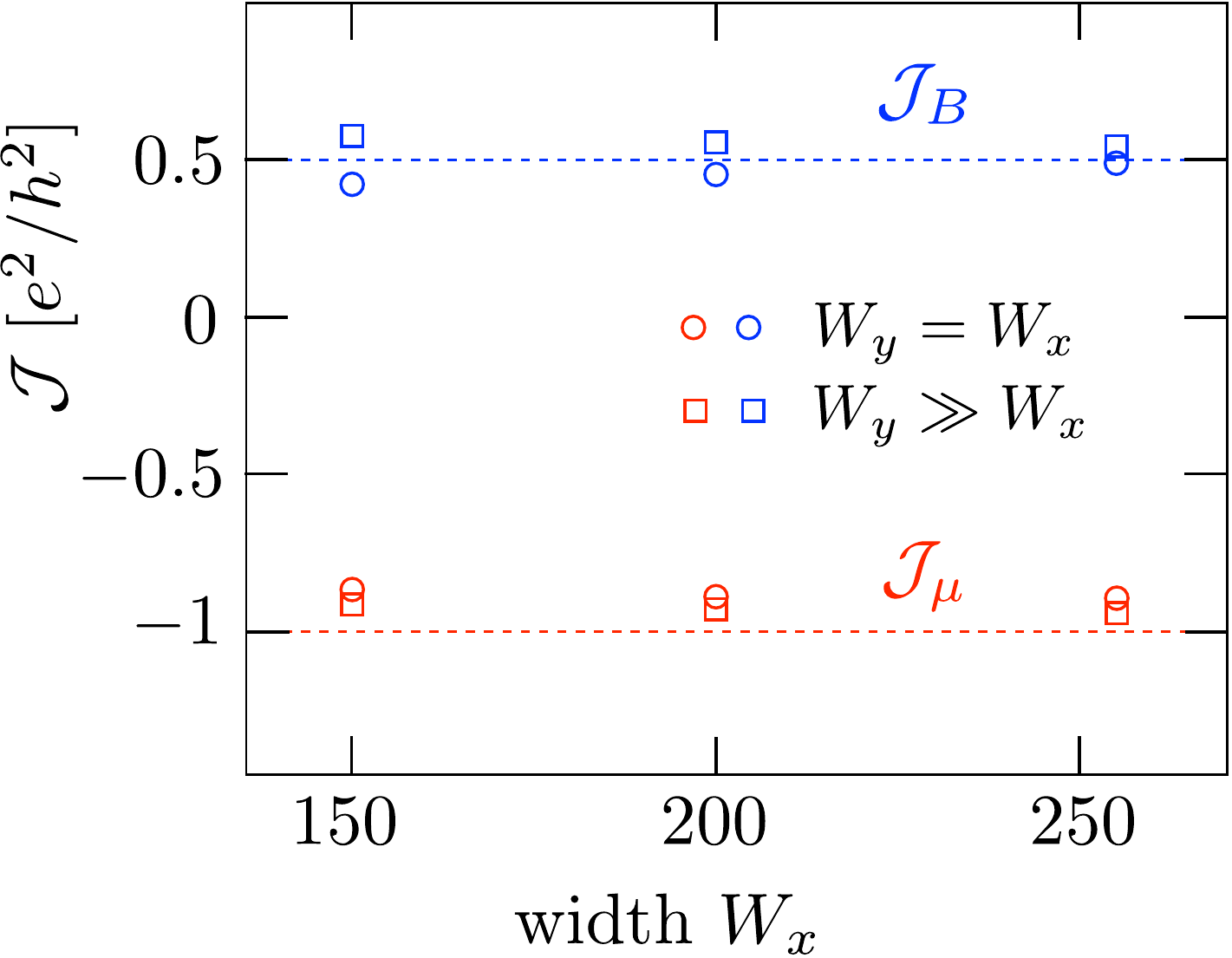}}
\caption{Results for ${\cal J}_B=\mu^{-1}\delta j/\delta B$ and
  ${\cal J}_\mu=B^{-1}\delta j/\delta\mu$ following from the
  scattering formula \eqref{IzeroT}, for the Weyl semimetal
  Hamiltonian \eqref{HMdef} with parameters as in Fig.\
  \ref{fig_bandstructure}. The data is shown at three different values
  of $W_x$, for two geometries: $W_y=W_x$ (circular symbols, hard-wall
  boundary conditions in both $x$ and $y$ directions) and
  $W_y=5000\gg W_x$ (square symbols, hard-wall boundary conditions along $x$, periodic
  boundary conditions along $y$).  }
\label{fig_results}
\end{figure}

We have calculated the current density $\delta j$ flowing along the wire in response to a slowly varying $\mu$ or $B$. In the former case we fix $B$ at $l_B=25$ and increase $\mu\equiv X$ from $X_0\equiv 0$ to $\delta X\equiv\delta\mu$, in the latter case we fix $\mu=0.196\,t$ and increase $B\equiv X$ from $X_0\equiv 0$ to $\delta X\equiv\delta B$. We obtain the CME coefficients in linear response,
\begin{equation}
{\cal J}_\mu\equiv B^{-1}\delta j/\delta\mu,\;\;{\cal J}_B=\mu^{-1}\delta j/\delta B,
\end{equation}
from the scattering formula \eqref{IzeroT}, with $T_n\equiv 1$ (no disorder, so unit transmission for all modes). The Fermi level is set at $E_{\rm F}=0$. Results are shown in Fig.\ \ref{fig_results}. We see that the numerical data points \cite{note2} lie close to the dashed lines given by
\begin{equation}
{\cal J}_\mu=-(e/h)^2,\;\;{\cal J}_B=\tfrac{1}{2}\times (e/h)^2.\label{JmuJBexpected}
\end{equation}

\begin{figure}[tb]
\centerline{\includegraphics[width=0.9\linewidth]{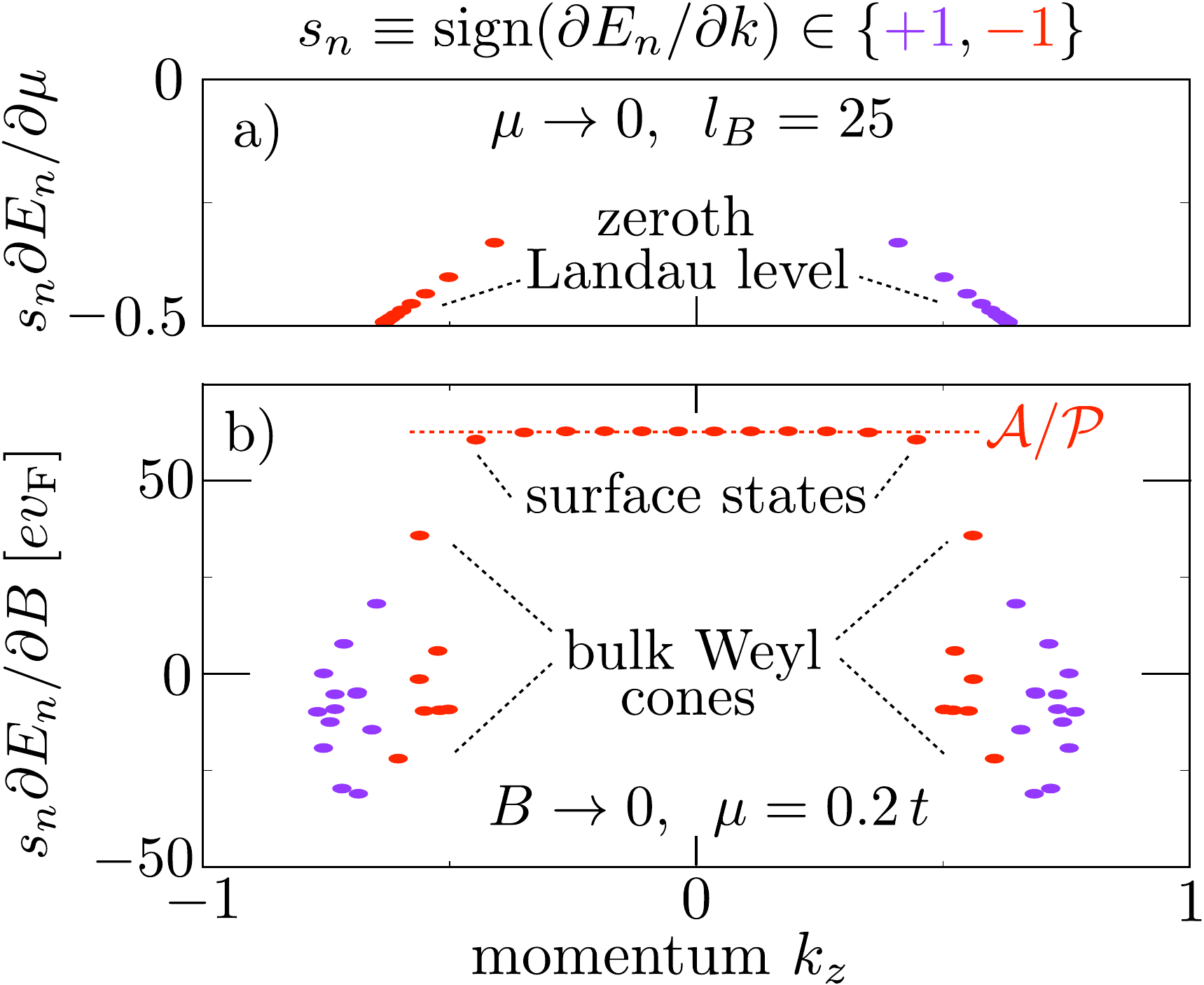}}
\caption{Contributions to ${\cal J}_\mu$ (panel a) and ${\cal J}_B$
  (panel b) from each individual mode, corresponding to the band
  structures shown in Figs.\ \ref{fig_bandstructure}b and
  \ref{fig_bandstructure}c. The sum of these contributions produces
  the total CME coefficient of Fig.\ \ref{fig_results}, at $W_y=W_x=255$. 
  The dotted line in panel b is the contribution
  \eqref{chinarc} expected from the Fermi arc Hamiltonian
  \eqref{Harcdef} for a surface enclosing an area ${\cal A}=(255)^2$
  with perimeter ${\cal P}=4\times 255$. The color of the data points
  distinguishes left-movers from right-movers,
  $s_n\equiv {\rm sign}\,(\partial E_n/\partial k)=+1$ (purple) or
  $-1$ (red).  }
\label{fig_details}
\end{figure}

The CME coefficient ${\cal J}_\mu$ agrees with the expected value from Eq.\ \eqref{jCME}, while the CME coefficient ${\cal J}_B$ has the opposite sign and is smaller by a factor of two. Inspection of the contributions from individual modes, plotted in Fig.\ \ref{fig_details}, indicates that surface states are behind the different response, as we now explain in some detail.

\subsection{Why surface Fermi arcs contribute to the magnetic response in the infinite-system limit}
\label{sec_Why}

Consider the propagating modes through a wire of diameter $W$. The number of surface modes scales $\propto W$, while the number of bulk modes scales $\propto W^2$, so one might surmise that surface contributions to the current density $I/W^2$ can be neglected in the limit $W\rightarrow\infty$. This is correct for ${\cal J}_\mu$ --- but not for ${\cal J}_B$, because each surface mode individually contributes an amount $\propto W$, so the total surface contribution scales $\propto W^2$, just like the bulk contribution.

To make this argument more precise, we consider the effective Hamiltonian of the surface Fermi arcs,
\begin{equation}
H_{\rm arc}=v_{\rm F} (p_s-e\Phi/{\cal P}) -v_{{\rm arc},z} p_z,\label{Harcdef}
\end{equation}
with $p_s$ the component of the momentum along the perimeter of the wire in the $x$-$y$ plane, of length ${\cal P}$ enclosing a flux $\Phi=B{\cal A}$ in an area ${\cal A}$. The energy spacing of the surface states at given momentum $p_z$ along the wire is $\delta E=hv_{\rm F}/{\cal P}$, so an energy separation $\mu$ of the Weyl cones pushes $N_{\rm arc}=\mu/\delta E=\mu{\cal P}/h v_{\rm F}$ surface modes through the Fermi level. Each contributes
\begin{equation}
\chi_n={\rm sign}(\partial E_n/\partial p_z)\times\partial E_n/\partial B=ev_{\rm F}{\cal A}/{\cal P}\label{chinarc}
\end{equation}
to the induced current, which is just its orbital magnetic moment. The total surface contribution takes on the universal value
\begin{equation}
{\cal J}_{B,{\rm arc}}=(e/h)N_{\rm arc}\chi_n/\mu{\cal A}=(e/h)^2.\label{JBarc}
\end{equation}
The red dotted line in Fig.\ \ref{fig_details}b confirms this reasoning.

\subsection{Bulk Weyl cone contribution to the magnetic response}

The numerical data in Fig.\ \ref{fig_results} indicates that the bulk Weyl cones contribute
\begin{equation}
{\cal J}_{B,{\rm bulk}}=-\tfrac{1}{2}(e/h)^2 \label{JBbulk}
\end{equation}
to the CME coefficient induced by a magnetic field, for a total ${\cal J}_{B,{\rm bulk}}+{\cal J}_{B,{\rm arc}}=\tfrac{1}{2}(e/h)^2$. We have not found a simple intuitive argument for Eq.\ \eqref{JBbulk}, but we do have  an explicit analytical calculation, see App.\ \ref{App_bulkcalculation}.

The difference between ${\cal J}_\mu$ and ${\cal J}_B$ goes against the original expectation \cite{Che13} that the low-frequency response to small variations in $\mu$ at fixed $B$ should be the same as to small variations in $B$ at fixed $\mu$. That there is no such reciprocity was found recently in two studies \cite{Ma15,Zho15} of currents induced by an oscillating magnetic field in an infinite isotropic system. Their bulk response has a different numerical coefficient than our Eq.\ \eqref{JBbulk} ($1/3$ instead of $1/2$), possibly because of the intrinsic anisotropy of a wire geometry.

\subsection{Interplay of surface Fermi arcs with bulk Landau levels}

\begin{figure}[tb]
\centerline{\includegraphics[width=1\linewidth]{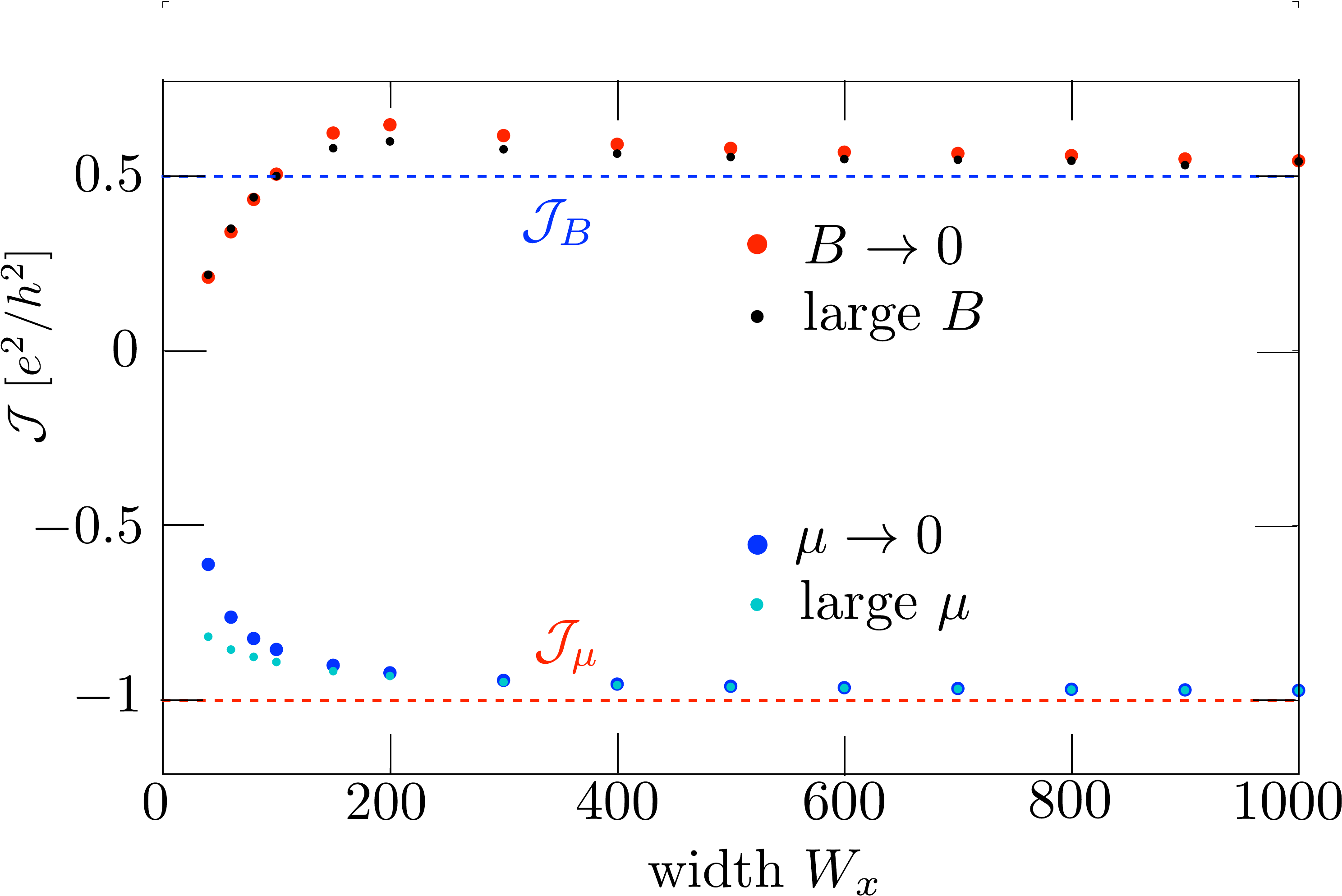}}
\caption{Same as Fig.\ \ref{fig_results}, but for a larger range of widths $W_x$ at fixed $W_y=5000$ (periodic boundary conditions in the $y$-direction). The data for ${\cal J}_B=\mu^{-1}\delta j(B)/\delta B$ is shown at $\gamma=0.1\,t\Rightarrow\mu=0.098\,t$ in the limit $B\rightarrow 0$ and at a large magnetic field in the Landau level regime ($l_B=50$). The data for ${\cal J}_\mu=B^{-1}\delta j(\mu)/\delta\mu$ is shown at $l_B=50$ in the limit $\mu\rightarrow 0$ and for a large $\mu=0.098\,t$.}
\label{fig_results2}
\end{figure}

\begin{figure}[tb]
\centerline{\includegraphics[width=0.9\linewidth]{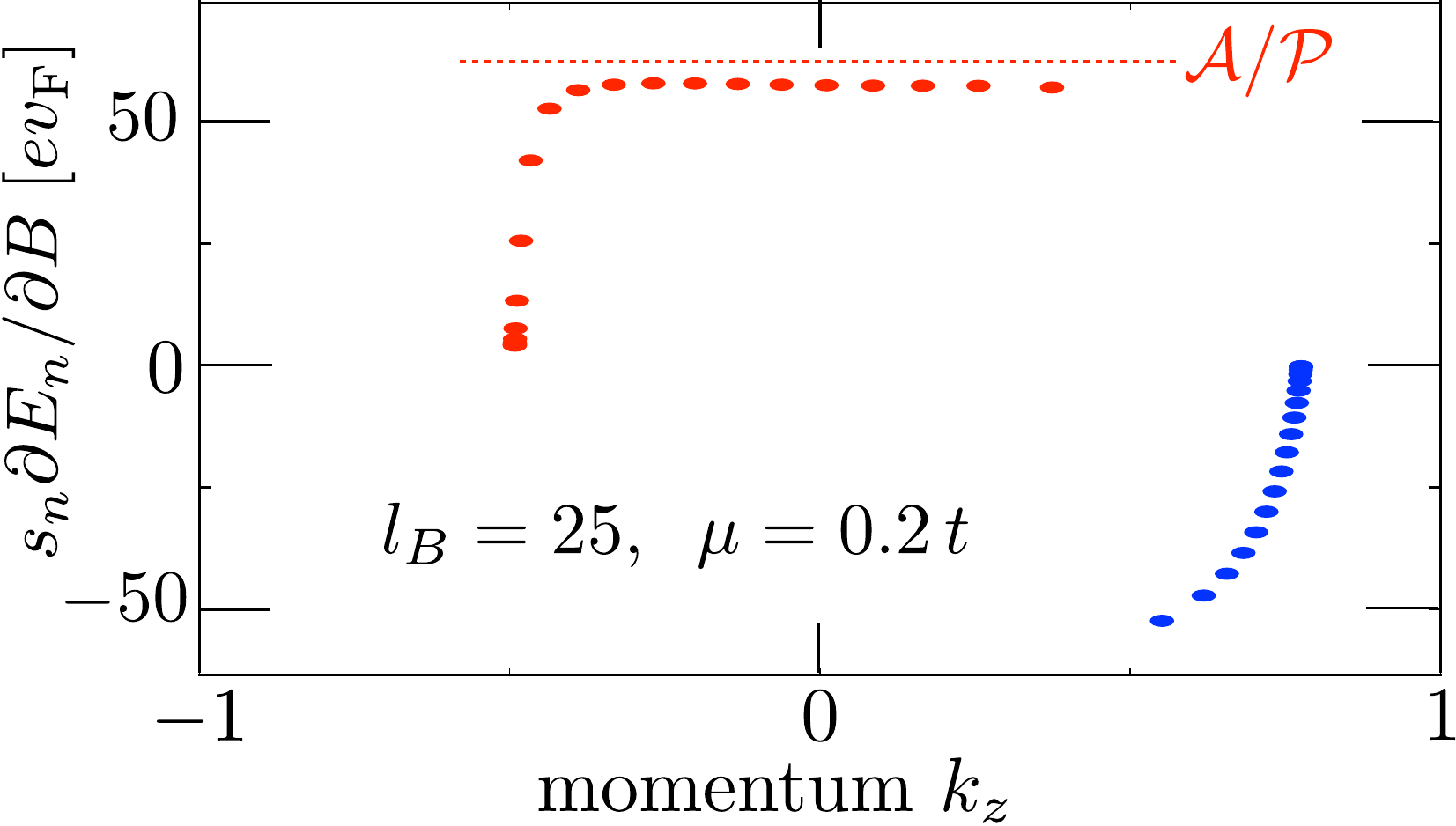}}
\caption{Same as Fig.\ \ref{fig_details}b, but now at a large magnetic field corresponding to the bandstructure in Fig.\ \ref{fig_bandstructure}d. The contribution from the surface Fermi arcs, close to ${\cal A}/{\cal P}$, goes to zero when they hybridize with the zeroth Landau level (for which $\partial E_n/\partial B=0$).
}
\label{fig_details2}
\end{figure}

So far we have considered the magnetic response in the zero-magnetic field limit, when the bulk contribution arises from Weyl cones. We can also ask for the current density $\delta j$ in response to a slow variation $\delta X\equiv\delta B$ around some nonzero $X_0\equiv B_0$, all at fixed $\mu$. As shown in Fig.\ \ref{fig_results2}, the magnetic response is the same whether we vary $B$ around zero or nonzero $B_0$. This is remarkable, because the bulk states are entirely different --- Weyl cones versus Landau levels, compare the band structures in Figs.\ \ref{fig_bandstructure}c and \ref{fig_bandstructure}d. The individual modes also contribute very differently to ${\cal J}_B$, compare Figs.\ \ref{fig_details}b and \ref{fig_details2}, and yet the net contribution is still close to $\frac{1}{2}\times (e/h)^2$. We have not succeeded in an analytical derivation of this numerical result.

\section{Finite-size effects}
\label{sec_finite}

We have seen that the surface Fermi arcs modify the magnetic response $\delta j/\delta B$ even in the limit that the size of the system tends to infinity. The response $\delta j/\delta\mu$ to an energy displacement $\mu$ of the Weyl cones is unaffected by the surface states in the infinite-system limit, given by Eq.\ \eqref{jCME} in that limit. There are however finite-size effects from the surface state contributions, which we consider in this section.

\begin{figure}[tb]
\centerline{\includegraphics[width=1\linewidth]{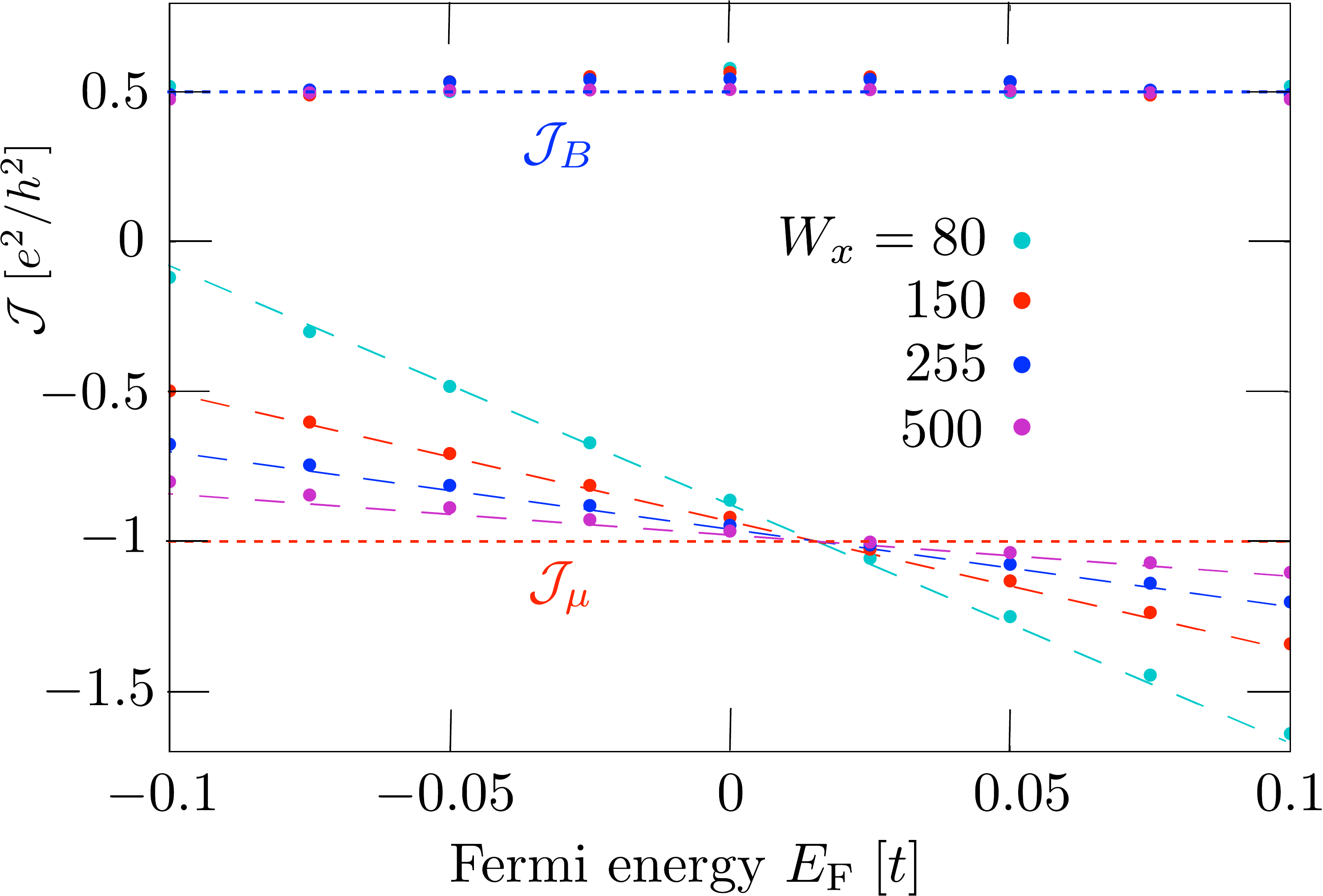}}
\caption{Fermi-energy dependence of the CME coefficients ${\cal J}_\mu=B^{-1}\delta j/\delta\mu$ (in the limit $\mu\rightarrow 0$ at $l_{B}=25$) and ${\cal J}_B=\mu^{-1}\delta j/\delta B$ (in the limit $B\rightarrow 0$ at $\mu=0.196\,t$), for different widths $W_x$ at fixed $W_y=5000$ (other parameters as in Fig.\ \ref{fig_bandstructure}, periodic boundary conditions in the $y$-direction). The horizontal dotted lines are the expected values \eqref{JmuJBexpected} in the limit of an infinite system, the dashed lines have a slope given by Eq.\ \eqref{JmufiniteEF}.
}
\label{fig_finiteEF}
\end{figure}

As shown in Fig.\ \ref{fig_finiteEF}, finite-size effects on ${\cal J}_\mu=B^{-1}\delta j/\delta\mu$ are sensitive to whether or not the Fermi level $E_{\rm F}$ is symmetrically arranged between the Weyl points ($E_{\rm F}=0$ in Fig.\ \ref{fig_bandstructure}). The earlier plots (Figs.\ \ref{fig_results} and \ref{fig_results2}) were for $E_{\rm F}=0$, when finite-size effects are small.  A variation of $E_{\rm F}$ away from the symmetry point has little effect on the magnetic response ${\cal J}_B=\mu^{-1}\delta j/\delta B$, provided $|E_{\rm F}|\lesssim|\mu|$. In contrast, the Fermi level displacement introduces a substantial size-dependence in ${\cal J}_\mu$.

Inspection of the band structure in Fig.\ \ref{fig_bandstructure}b shows that the degeneracy of the zeroth Landau level increases with increasing $E_{\rm F}$, because surface modes are converted into bulk modes, at a rate given by the inverse of the level spacing $\delta E=h v_{\rm F}/{\cal P}$ (cf.\ Sec.\ \ref{sec_Why}). Each bulk mode contributes $-(e/h)\delta\mu/{\cal A}$ to the induced current density $\delta j$, so we expect a finite-size correction to ${\cal J}_\mu=B^{-1}\delta j/\delta\mu$ equal to $-(e/h)(B{\cal A})^{-1} (E_{\rm F}/\delta E)$, hence
\begin{equation}
{\cal J}_\mu=-(e/h)^2\left(1+\frac{\cal P}{\cal A}\frac{E_{\rm F}}{eBv_{\rm F}}\right).\label{JmufiniteEF}
\end{equation}
As seen in Fig.\ \ref{fig_finiteEF}, the slope of the $E_{\rm F}$ dependence of $J_{\mu}$ is accurately described by Eq.\ \eqref{JmufiniteEF}.

\section{Conclusion and discussion of disorder effects}
\label{sec_conclude}

Fig.\ \ref{fig_results2} summarizes our main finding: It is known \cite{Che13,Gos13,Cha14,Ala15,Ma15,Zho15} that the chiral magnetic effect in a Weyl semimetal can be driven either by a slowly varying inversion-symmetry breaking $\mu$ or by a slowly varying magnetic field $B$. Contrary to the expectation from an infinite system, we find for a finite system that the induced current in the two cases has opposite sign. The difference is due to the surface Fermi arcs, but it is not a finite-size effect: The surface modes and the bulk modes give comparable contribution to the magnetic response no matter how large the system is, because the smaller number of surface modes is compensated by their stronger $B$-sensitivity.

This finding results from a scattering formulation of the chiral magnetic effect, that we have developed as an alternative to the established Kubo formulation \cite{Kha09,Lan11}. Similarly to the Landauer formula for electrical conduction, the scattering formula \eqref{IzeroT} is ideally suited to describe finite and disordered systems, without translational invariance. Here we focused on the surface effects in a finite system, but in closing we briefly consider the disorder effects.

A qualitative prediction can be made without any calculation. In Eq.\ \eqref{IzeroT} disorder reduces the contribution from each mode $n$ by its transmission probability $T_n$. Assume that the disorder potential is smooth on the scale of the lattice constant $a$, so that it predominantly couples nearby modes in the Brillouin zone (with $k_n$'s differing by much less than $1/a$). This coupling can only lead to backscattering, reducing $T_n$ below unity, if it involves both left-moving and right-moving modes. 
Inspection of Fig.\ \ref{fig_details}b shows that the surface modes are insensitive to backscattering, because they all move in the same direction along the wire, in contrast to the bulk Weyl cones. Disorder will therefore reduce the Weyl cone contribution ${\cal J}_{B,{\rm bulk}}=-\frac{1}{2}(e/h)^2$ to the magnetic response, without affecting the arc state contribution ${\cal J}_{B,{\rm arc}}=(e/h)^2$. Since these contributions have opposite sign, we predict that disorder will \textit{increase} the magnetically induced current.

For sufficiently strong disorder the bulk contribution to ${\cal J}_B$ may be fully suppressed, leaving a $B$-induced current density equal to $j=(e/h)^2\mu B$, carried entirely by the surface Fermi arc. This has the same topological origin as the zeroth Landau level that carries the $\mu$-induced current \eqref{jCME} --- both the Fermi arc and the zeroth Landau level connect Weyl cones of opposite chirality \cite{Wan11,Hal14}.  It has been argued \cite{Ma15,Zho15} that the chiral magnetic effect produced by an oscillating $B$ is fundamentally different from that produced by an oscillating $\mu$, because the former lacks the topological protection that is the hallmark of the latter. By including surface conduction we can now offer an alternative perspective: Both the $\mu$-response and the $B$-response are similarly protected by chirality, the difference is that one is a bulk current and the other a surface current.  

From an experimental point of view, the inversion-symmetry breaking that sets $\mu$ is hardly adjustable, preventing a direct measurement of $\delta j/\delta \mu$, while the magnetic field induced current $\delta j/\delta B$ seems readily accessible. We note that Landau levels are not required for the $B$-response, so one can work with a nanowire of width small compared to the magnetic length. In such a quasi-one-dimensional system long-range impurity scattering may localize the bulk states, without significantly affecting the chiral surface states. One would be searching for an oscillating current $I(\omega)\cos\omega t$ along the wire in response to an oscillating parallel magnetic field. The frequency $\omega$ should be below $\mu$ and above the inelastic relaxation rate of the surface modes. The magnetic response is quasi-{\sc dc}, showing a plateau in this $\omega$-range that would distinguish it from any electrically induced {\sc ac} current $I(\omega)\propto\omega$.

A final word on nomenclature. The non-topological bulk contribution to the $B$-induced current has been termed the ``gyrotropic magnetic effect'' \cite{Zho15}. Because the $B$-induced current in the surface Fermi arc originates from the same chiral anomaly as the $\mu$-induced current in the zeroth Landau level, we use the name ``chiral magnetic effect'' for both \cite{name}.

\acknowledgments

We have benefited from discussions with I. Adagideli, T. E. O'Brien, V. V. Cheianov, and B. Tarasinski. This research was supported by the Foundation for Fundamental Research on Matter (FOM), the Netherlands Organization for Scientific Research (NWO/OCW), and an ERC Synergy Grant.

\appendix

\section{Analytical calculation of the bulk contribution to the magnetic response}
\label{App_bulkcalculation}

We wish to derive the result \eqref{JBbulk} for the contribution 
\begin{equation}
\mu^{-1}\delta j_{\rm bulk}/\delta B=-\tfrac{1}{2}(e/h)^2\label{deltajbulkWeyl}
\end{equation}
from the bulk Weyl cones to the magnetic response. We assume that the two Weyl cones are non-overlapping at the Fermi energy (as they are in Fig.\ \ref{fig_bandstructure}c), so we can consider them separately. 

A single Weyl cone has Hamiltonian
\begin{equation}
H_{\rm Weyl}= v_x k_x\sigma_x+ v_y k_y\sigma_y+ v_z k_z\sigma_z.\label{HWeyl}
\end{equation}
(We have set $\hbar\equiv 1$ for ease of notation, but we will reinstate it at the end.) For generality, we allow for an anisotropic velocity $(v_x,v_y,v_z)$. The modes propagating along the cylindrical wire have energy $E_{n}(k_z)$. We seek the magnetic moment $\partial E_{n}(k_z)/\partial B$ for an infinitesimal magnetic field $B$ in the $z$-direction (along the axis of the cylinder).

For sufficiently large transverse dimensions $W_x,W_y$ the boundary conditions should be irrelevant for the bulk response, and we use this freedom to simplify the calculation. To isolate the bulk contribution we prefer a boundary condition that does not bind a surface state. 

In the $y$-direction we impose periodicity, so that $k_y$ is a good quantum number. The system is then represented by a hollow cylinder of circumference $W_y$, with an inner and an outer surface at $x=0$ and $x=W_x$. We can use periodic or antiperiodic boundary conditions, 
\begin{equation}
\begin{split}
&k_y=2\pi n/W_y\;\;{\rm or}\;\;k_y=2\pi (n+\tfrac{1}{2})/W_y,\\
&\qquad\qquad n=0,\pm 1,\pm 2,\ldots,
\end{split}\label{ybc}
\end{equation}
in the large-$W_y$ limit it makes no difference.

In the $x$-direction we choose a zero-current boundary condition. A simple choice is to take the spinor $\psi(x,y)$ as an eigenfunction of $\sigma_y$ at the two surfaces $x=0$ and $x=W_x$, 
\begin{equation}
\lim_{x\rightarrow 0}\psi(x,y)=f(y)\begin{pmatrix}
1\\
i
\end{pmatrix},\;\;\lim_{x\rightarrow W_x}\psi(x,y)=g(y)\begin{pmatrix}
1\\
i
\end{pmatrix},\label{boundarycond}
\end{equation}
for arbitrary complex functions $f(y)$, $g(y)$. This boundary condition corresponds to confinement by a mass term $\propto\sigma_z$ of infinite magnitude and opposite sign at the two surfaces. No surface state is produced by mass confinement. For $k_z=0$ the sign change of the mass term does produce a spurious chiral state $\psi=e^{ik_y y}{1\choose i}$, $E=v_y k_y$, which carries no current in the $z$-direction and can therefore be ignored.

\begin{widetext}
The solution of the eigenvalue equation $H_{\rm Weyl}\psi=E\psi$ that satisfies the boundary condition \eqref{boundarycond} is given by
\begin{align}
\psi(x,y)={}&\frac{1}{Z}e^{ik_y y}\left[(E-i v_x k_x- v_yk_{y}+ v_z k_{z})\begin{pmatrix}E+ v_z k_{z}\\
 v_x k_x+i v_yk_{y}
\end{pmatrix}e^{i  k_x x}\right.\nonumber\\
&\qquad\quad-\left.(E+i v_x k_x- v_y k_{y}+ v_z k_z)\begin{pmatrix}E+ v_z k_{z}\\
 -v_x k_x+i v_y k_{y}
\end{pmatrix}e^{-ik_x x}\right],\label{eq:bulk}
\end{align}
\begin{equation}
k_x=\pi m/W_x,\;\;m=1, 2, 3,\ldots.\label{xbc}
\end{equation}
\end{widetext}
The band structure $E_{nm}(k_z)$ is determined by the dispersion relation
\begin{equation}
E^2=(\pi mv_x/W_x)^2+(2\pi n v_y/W_y)^2+(v_z k_z)^2.\label{dispersion}
\end{equation}
Normalization $\langle\psi|\psi\rangle=1$ gives
\begin{equation}
Z^2=8W_x W_y E(E-v_y k_y)(E+v_z k_z)^2.
\end{equation}

\begin{figure}[tb]
\centerline{\includegraphics[width=0.9\linewidth]{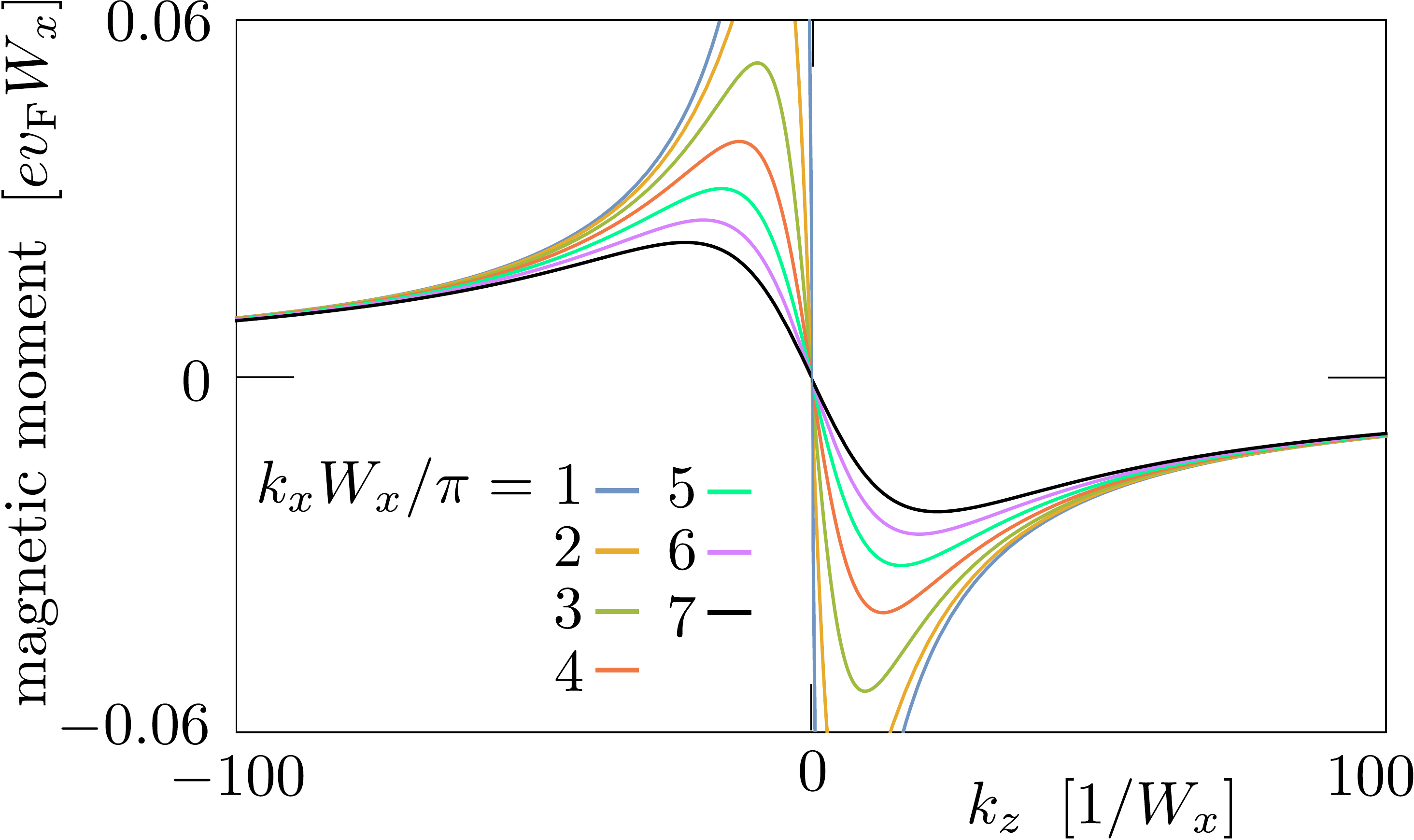}}
\caption{Magnetic moment \eqref{magneticmoment} of a single Weyl cone (isotropic, $v_x=v_y=v_z\equiv v_{\rm F}$), summed over $+k_y$ and $-k_y$, as a function of $k_z$ for the seven lowest quantized values of $k_x=m\pi/W_x$. The quantization of $k_y$ can be ignored for $W_y\gg W_x$, so the discrete modes merge into a continuous curve. 
}
\label{fig_magneticmoment}
\end{figure}

The magnetic response is induced by the vector potential $A_y=B(x+X_0)$, with an offset $X_0$ that accounts for the flux $BW_y X_0$ enclosed by the inner surface of the cylinder. The magnetic moment results from
\begin{align}
\frac{\partial E}{\partial B}&=\langle \psi| \partial H/\partial B|\psi\rangle=-ev_y\langle \psi|  (x+X_0)\sigma_y|\psi\rangle\nonumber\\
&=-\frac{ev_x v_y v_z k_z}{2E(E-v_y k_y)}-\frac{ev_y^2 k_y}{E}(X_0+W/2).\label{dEdBresult}
\end{align}
The second term is the magnetic moment of a charge $e$ circulating along the inner surface of the cylinder with velocity $\partial E/\partial k_y=v_y^2 k_y/E$. It drops out when we sum the contributions from $+k_y$ and $-k_y$, producing the magnetic moment
\begin{equation}
\sum_{\pm k_y}\frac{\partial E}{\partial B}=-\frac{ev_x v_y v_z k_z}{E^2-v_y^2 k_y^2}=-\frac{ev_x v_y v_z k_z}{v_x^2 k_x^2+v_z^2 k_z^2}\label{magneticmoment}
\end{equation}
plotted in Fig.\ \ref{fig_magneticmoment}.

We fix the energy $E_{nm}=E$, adjusting $k_z$ accordingly for each $n$ and $m$. Both $+k_z$ and $-k_z$ satisfy the dispersion relation \eqref{dispersion}. We consider separately the sum over the magnetic moment of the modes with $k_z>0$ and $k_z<0$, so that we can distinguish left-movers from right-movers in the scattering formula \eqref{IzeroT}. In the large-$W$ limit the sum over $k_x$ and $k_y$, quantized by Eqs.\ \eqref{ybc} and \eqref{xbc}, can be replaced by an integration over the $k_x$-$k_y$ plane,
\begin{equation}
\sum_{nm}{}'\mapsto \frac{W_x W_y}{2\pi^2}\int_{0}^\infty dk_x\int_{-\infty}^\infty dk_y\, \theta(E^2-v_x^2 k_x^2-v_y^2 k_y^2),
\end{equation}
with $\theta(s)$ the unit step function. The prime in the summation indicates that we include only half of the modes, with a given sign of $k_z$. The integral over Eq.\ \eqref{dEdBresult} is readily evaluated in polar coordinates,
\begin{align}
\sum_{nm}{}'\,\left.\frac{\partial E_{nm}}{\partial B}\right|_{E_{nm}=E}&=-({\rm sign}\,k_z)\frac{eW_x W_y}{4\pi^2 E}\nonumber\\
&\times\int_{-\pi/2}^{\pi/2} d\phi\int_0^{|E|} rdr\,\frac{\sqrt{E^2-r^2}}{E-r\sin\phi}\nonumber\\
&=-({\rm sign}\,k_z)\frac{eW_x W_y |E|}{8\pi}.
\end{align}

The quantity $\chi_{nm}$ that determines the magnetically induced current $\delta I/\delta B$ according to Eq.\ \eqref{IzeroT} is the magnetic moment $\partial E_{nm}/\partial B$ times the sign of the velocity $\partial E_{nm}/\partial k_z$ in the $z$-direction. The sign of the velocity in a single Weyl cone (with Weyl point at $\bm{k}=0$, $E=0$) equals the product of the sign of $k_z$ and the sign of $E_{nm}$, hence
\begin{align}
\sum_{nm}\chi_{nm}&=\sum_{nm}\left({\rm sign}\,\frac{\partial E_{nm}}{\partial k_z}\right)\left.\frac{\partial E_{nm}}{\partial B}\right|_{E_{nm}=E}\nonumber\\
&=-\frac{eW_x W_y E}{4\pi\hbar}.
\end{align}
In the last equation we have reinstated the $\hbar$ that we had previously set to unity. There is no prime in the summation because now both signs of $k_z$ are included. 

We conclude that the contribution to the induced current density $\delta j=\delta I/W_xW_y$ from a single Weyl cone at energy $E=\mu/2$ is
\begin{equation}
\delta j=\frac{e}{h}\frac{\delta B}{W_x W_y}\sum_{nm}\chi_{nm}=-\tfrac{1}{4}(e/h)^2 \mu\delta B.
\end{equation}
The other Weyl cone contributes the same amount, for a total CME coefficient given by Eq.\ \eqref{deltajbulkWeyl}.


\begin{thebibliography}{99}
\bibitem{note1} To avoid a confusion of minus signs, we assign charge $+e$ to the carriers. The final result for the induced current contains $e^2$, so this sign convention does not affect it.
\bibitem{note2} The derivative $\partial E_n/\partial X$ needed in the scattering formula \eqref{IzeroT} can be calculated most easily from the Hellmann-Feynman equation $\partial E_n/\partial X=\langle\psi_n|\partial H/\partial X|\psi_n\rangle$, with $\psi_n$ the eigenfunction of mode $n$. In this way a numerical differentiation of the energy spectrum can be avoided.
\bibitem{name}  We cannot resist mentioning the name suggested to us by Roger Mong: Since the chiral anomaly (imbalance of left-movers and right-movers) in a surface Fermi arc does not require Landau levels, one might call it the ``chiral anomalous anomaly'', by analogy with the name ``quantum anomalous Hall effect'' for the quantum Hall effect without Landau levels.
\end{thebibliography}
\end{document}